\documentclass[11pt]{article}
\pdfoutput=1
\title{\textbf{A sound nebula: the origin of the Solar System in the field of a standing sound wave}}

\author{S. Beck\textsuperscript{1}, V. Beck\textsuperscript{1,*}  \\ \\
 \textsuperscript{1}Alfa-Carbon\textsuperscript{\textregistered} Beck, Berlin, Germany \\
 \textsuperscript{*}Correspondence: v.beck@alfa-carbon.de \\
 }
\date{}

\usepackage{graphicx}

\usepackage[font=small,labelfont=bf]{caption}

\usepackage{geometry}
\usepackage{hyperref}

\begin{document}

\maketitle

\begin{abstract}
According to the planetary origin conceptual model proposed in this paper, the protosun centre of the pre-solar nebula exploded, resulting in a shock wave that passed through it and then returned to the centre, generating a new explosion and shock wave. Recurrent explosions in the nebula resulted in a spherical standing sound wave, whose antinodes concentrated dust into rotating rings that transformed into planets. The extremely small angular momentum of the Sun and the tilt of its equatorial plane were caused by the asymmetry of the first, most powerful explosion. Differences between inner and outer planets are explained by the migration of solid matter, while the Oort cloud is explained by the division of the pre-solar nebula into a spherical internal nebula and an expanding spherical shell of gas. The proposed conceptual model can also explain the origin and evolution of exoplanetary systems and may be of use in searching for new planets.
\end{abstract}

\section{Introduction}

\paragraph{}The classical theory of the origin of the Solar System is based on the Kant -- Laplace nebular hypothesis \cite{kant1755, lapl1796} suggesting that the Sun and the planets condensed out of a spinning nebula of gas and dust. During the condensation process, the spin of the nebula accelerated (the 'pirouette' effect), and the resulting distribution of angular momentum caused it to form a disc. The center of the nebula evolved into a hot, highly compressed gas region -- the protosun. Concentration and coalescence of dust particles in the remainder of the spinning disc led to the formation of planets orbiting the Sun. 

This theory accounts for the general nature of the origin of the Solar System, but it cannot explain many of the observed facts. One problem is the angular momentum distribution: the Sun has more than 99.8\% of the entire system mass, but only about 0.5\% of the total angular momentum, with the remaining 99.5\% residing in the orbiting planets. The hypothesis also fails to explain the 7$^\circ$ tilt of the Sun's equatorial plane relative to the average orbital plane of the planets. Another serious problem is the distinction between small solid-surface inner planets and outer gas giants. Moreover, the observed regularity in planetary distances from the Sun -- the so-called Titius-Bode law -- has no explanation. 

It is difficult to explain the existence of the Oort cloud beyond the Solar System's planets. It consists of trillions of small objects composed of dust and water, ammonia and methane ice and it is believed that these objects were scattered outwards by the gas giants at the planetary formation stage and then acquired distant circular orbits (out to about one light year) as a result of gravitational forces due to nearby stars. Such an Oort cloud emergence scenario seems very unlikely for such a large number of bodies. 

Neptune is the most distant gas planet. Based on the decreasing series of giant planet masses – Jupiter: 318 earth masses (M$_\oplus$); Saturn: 95.3 M$_\oplus$; Uranus: 14.5 M$_\oplus$ – we could expect Neptune's mass to be several times smaller than that of Uranus. Such a mass distribution can be explained by decreasing density of the gas nebula from the centre to the periphery, so that each planet has less gaseous substance than the previous one. In reality, Neptune has a mass of 17.5 M$_\oplus$, which is greater than the mass of Uranus. 

Authors, such as for example, Chamberlin \cite{cham1901}, Moulton \cite{moul1905} , Schmidt \cite{schm1944}, von Weizsäcker \cite{weiz1944}, McCrea \cite{mccr1960}, Woolfson \cite{wool1964} and Safronov \cite{safronov1972} have offered a variety of scenarios for the Solar System's formation. However, none of the existing theories is able to give a comprehensive picture of the emergence and development of the planetary system. A large number of recently discovered exoplanets (1955 confirmed planets by February 2016, NASA Exoplanet Archive\footnote{http://exoplanetarchive.ipac.caltech.edu}) allows us to estimate the number of planets in our galaxy as many billions, so that a comprehensive understanding of the processes leading to the emergence of planetary systems becomes critically important. In this article, we present a new conceptual model of the Solar System's formation from a pre-solar gas and dust nebula in the field of a standing sound wave.

\section{Theoretical construction}

\subsection{A spherical standing sound wave}

\paragraph{}It is generally accepted that during a sufficiently strong process of heating, gravitational contraction in the centre of the pre-solar nebula started thermonuclear fusion of hydrogen into helium. A large amount of energy was emitted and radiation pressure prevented the further contraction of the gas nebula. However, thermonuclear fusion could not start quietly; the process resulted in an explosion that caused a spherical shock wave originating from the central region. The specific explosion mechanism accompanied by the thermonuclear fusion processes is beyond the scope of this article; we note here only that the power of the explosion would have been large enough for the shock wave to spread all over the pre-solar nebula\footnote{Appendix A outlines some arguments that allow us to make some assumptions about the evolution of protostars.}.  A long time observed active processes, such as jets and outflows associated with star formation, e.g. \cite{ball1995}-\cite{stan2003}, suggests that the explosions fundamentally inherent in the birth of stars and can start the emergence of planetary systems as described in our model.

As the wave propagated, the gas particles in the pre-solar nebula oscillated radially, for two reasons: first, the difference in gas pressure; and secondly, the gravitational pull toward the centre of the nebula. The second factor began to play a crucial role at large distances from the central attracting mass concentrations: at some point, the accelerated gas particles at the wave front would no longer return under the influence of gravity. Gas density at this distance from the centre would have been so low that the pressure difference could no longer cause the return movement of gas particles. The peripheral part separated, and the pre-solar nebula was thus divided into a central spherical region and an expanding spherical shell of gas (Figure 1). 

\begin{figure}[ht]
\centering
\includegraphics[width=0.6\textwidth]{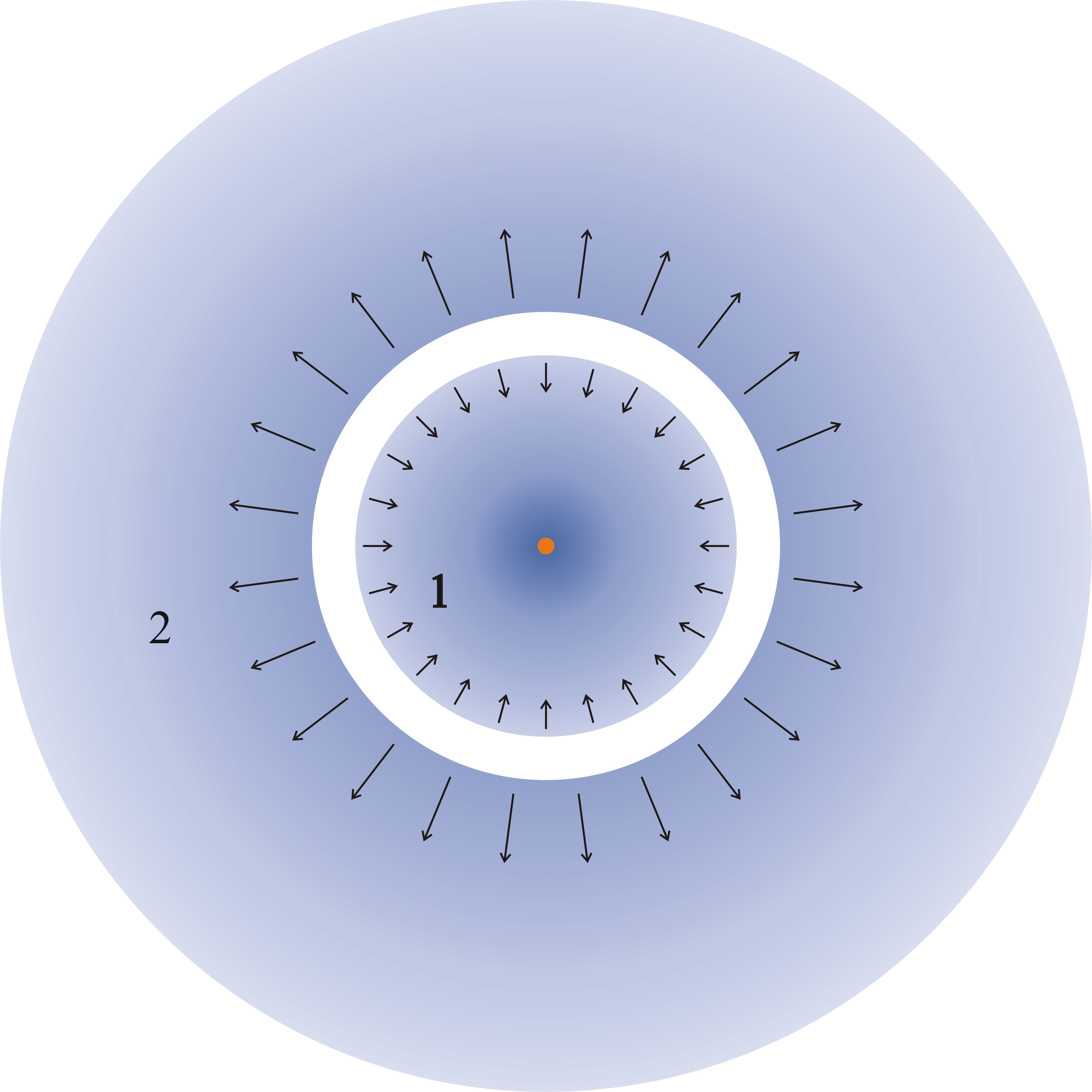}
\caption{Separation of the pre-solar nebula into (1) a central spherical gas nebula and (2) an expanding spherical shell.}
\end{figure}

Gas particles at the boundary of the central spherical nebula fell under gravity towards the centre then stopped when the pressure of the lower gas layers exceeded the gravitational pull, and began to move in the opposite direction. As a result of the interaction between gravity and contracting gas back-pressure, the boundary of the spherical nebula began to oscillate radially and a reverse spherical wave propagated from the periphery towards the centre. The sound wave would have appeared to reflect from the gravitational pause -- a spherical shell where the gas particles velocities were not high enough to overcome the gravitational pull from the centre of the nebula and the gas particles that had moved outwards returned due to oscillations; that is, their velocities were lower than the escape velocity:
\begin{equation}
\nu_e=\sqrt{\frac{2GM}{r}} ,
\end{equation}
where G is the gravitational constant, M is the mass of the attraction centre and r is the distance of the gravitational pause from the centre.

The centre of the pre-solar nebula became quiescent after the first massive explosion: the central part expanded, since the gravitational field was not yet intense enough to resist the much increased gas pressure, and the thermonuclear fusion of hydrogen combustion diminished. A large mass of gas was ejected into the surrounding space while the compact region of compressed and heated gas -- the protosun -- remained at the nebula's core. Several hundred years after this, the wave returning from the boundary of the spherical nebula reached the protosun, concentrated in the centre, and then began to propagate towards the periphery again: a rapid pressure increase resulted in a dramatic rise in temperature at the centre of the protosun, generating a new hydrogen explosion, much weaker than the first one but still strong enough to give extra energy to the wave reflected from the central region. The reflected wave travelled all the way from the centre to the periphery of the spherical nebula, was reflected from the gravitational pause and then returned to the centre again, causing another explosion. This process, repeated several times, eventually established regular oscillations: the wave propagated from the centre to the edge, was reflected from the boundary of the spherical nebula, and returned, causing another explosion that compensated for wave energy loss. Thus, the wave caused explosions while acquiring the energy it needed, establishing a self-sustaining process whose period, defined by the free oscillations of the spherical nebula boundaries, equalled many tens of years.

The acoustic radiation pressure prevented gravitational contraction of the pre-solar nebula and compensated for deviations from the spherical shape\footnote{Here and below, for simplicity, we are talking about a spherical shape. Needless to say, a rotating nebula takes the form of an ellipsoid of revolution; it can be shown that all the derivations for spherical sound waves retain also for the rotating nebula in the form of an ellipsoid.}  that might result, for example, from the gas turbulence. Waves travelling from the centre and from the periphery interfered with each other to form a giant spherical acoustic cavity resonator the size of the modern Solar System, including the Kuiper belt and the scattered-disc (Figure 2). This resonator contained a standing sound wave with nodes and antinodes, the number of which would have been at least 10 for the pre-solar system: 8 planets + the asteroid belt + the scattered-disc region. 

\begin{figure}[ht]
\centering
\includegraphics[width=0.6\textwidth]{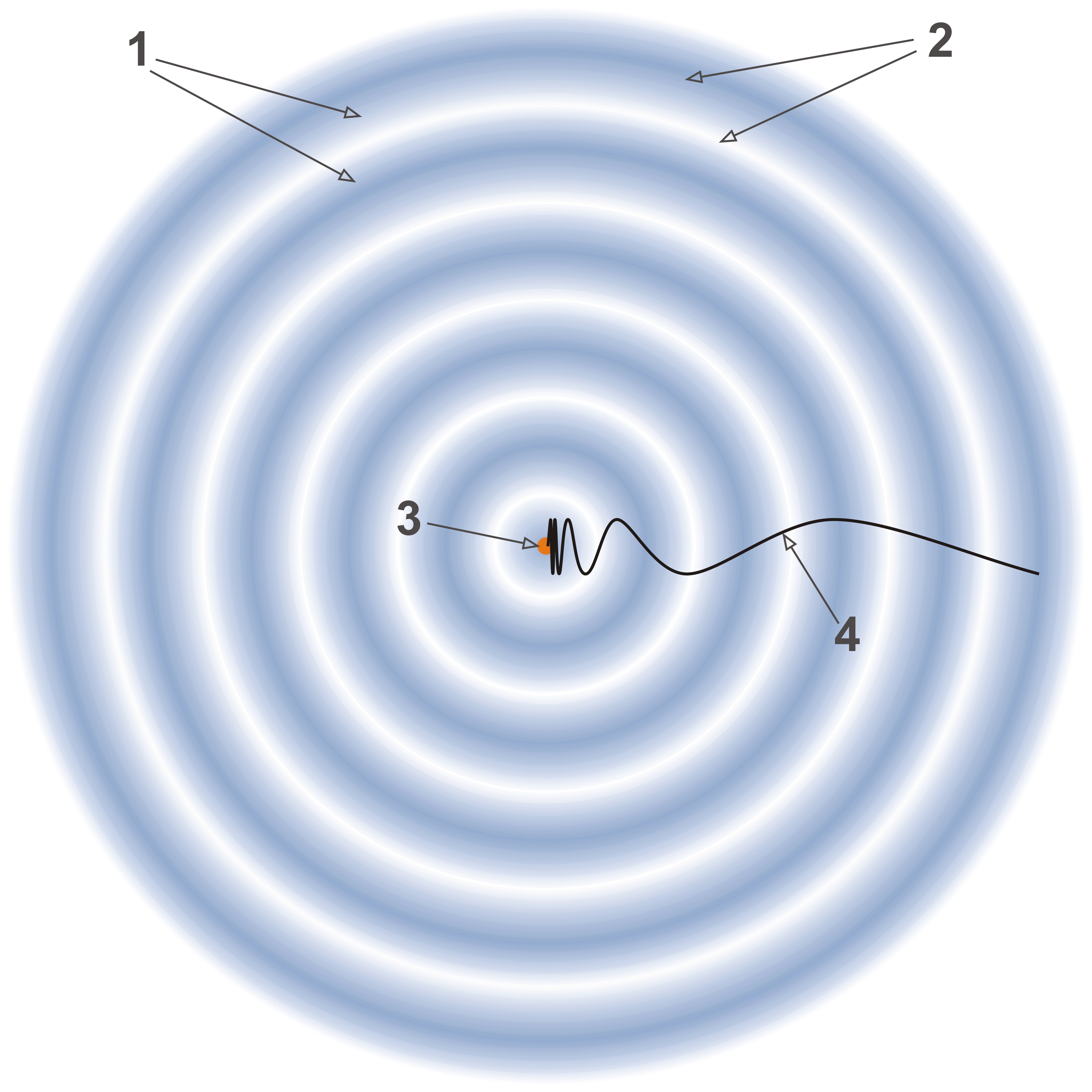}
\caption{The standing wave in the pre-solar nebula: (1) nodes, (2) antinodes (gas compression and expansion), (3) protosun, (4) real-scale wave. Nodes and antinodes are evenly spaced for illustration purposes; in reality, separations would increase dramatically with increasing distance from the centre.}
\end{figure}

This article does not include a calculation of exact distances between the antinodes of the standing wave; we may simply state that the wavelength increases with increasing distance from the centre of attraction. Section [3.4] outlines some regularities that allow us to make some observations about the physical reasons underlying planetary positions in the Solar System.

\subsection{Dust concentration in the antinodes}

\paragraph{}In our model, the pre-solar system can keep 'sounding' for many millions of years, as the periodic central explosions significantly retard the gravitational contraction of the protosun and the acoustic radiation pressure stabilizes the gas nebula, preventing its collapse. The dust present in the pre-solar nebula gradually concentrates in the antinodes of the acoustic oscillating system. Apart from gas viscosity, the process of solid particle concentration in the standing wave also relies on attraction from the large gas masses that periodically emerge in the antinodes of the standing wave. This attractive force makes the dust particles move towards the gas clusters in the antinodes and collide with each other, causing redistribution of their velocity vectors in such a way that the dust particles come to rest in the centre of the antinode (Figure 3). 

\begin{figure}[ht]
\centering
\includegraphics[width=0.6\textwidth]{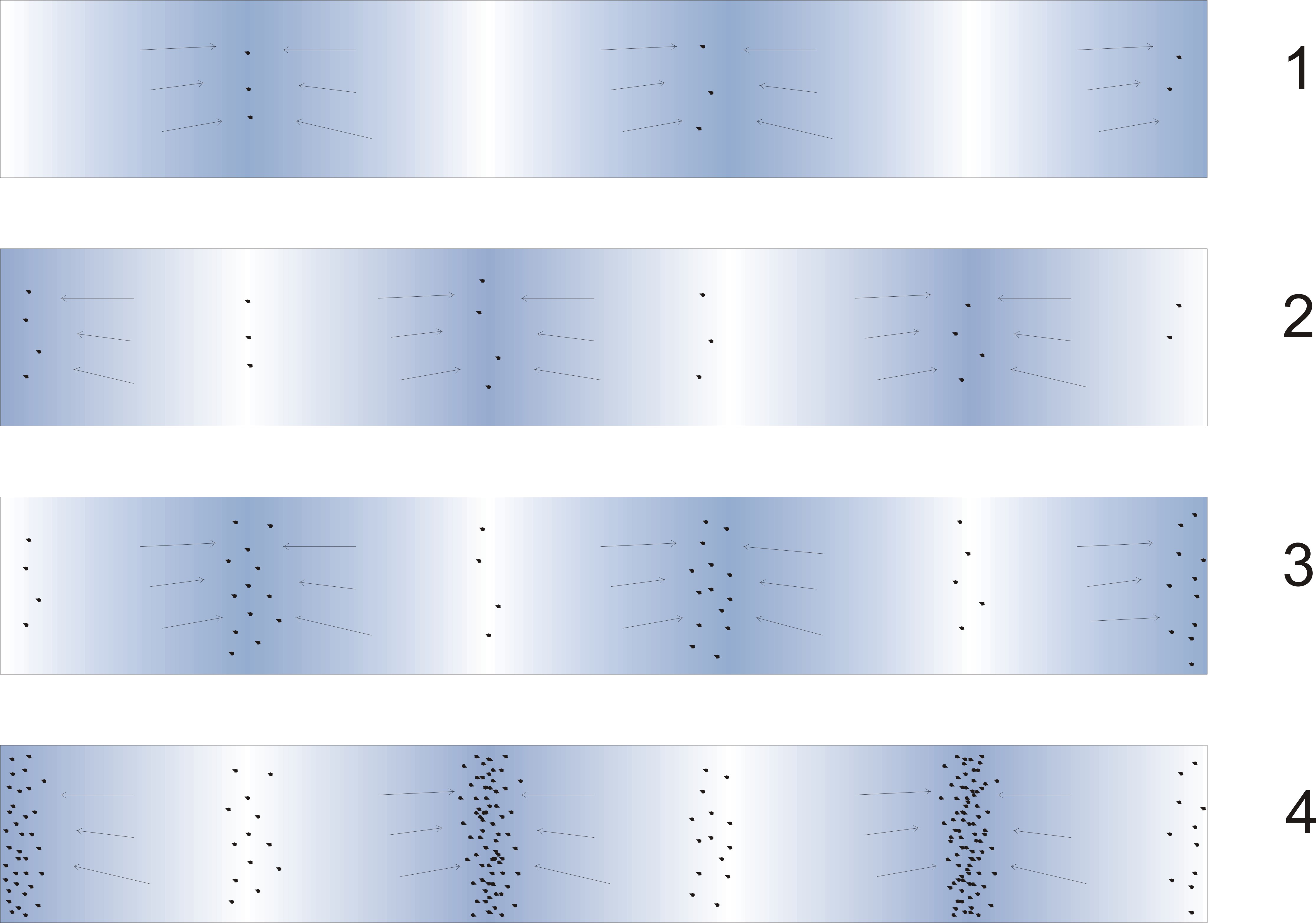}
\caption{Dust concentration in the antinodes of the standing wave during four successive phases of compression and expansion. Dark areas stand for gas compression in the antinodes, light areas for expansion.}
\end{figure}

During the gas expansion phase, dust particles in each antinode are gravitationally attracted towards two neighbouring antinodes. The two gravitational sources in these antinodes largely cancel each other out and, moreover, are remote, so that the dust particles in the gas expansion phase mostly remain in the same place. The next phase of compression attracts new dust particles to the antinode, which also come to rest due to collisions with each other and to the viscosity of the compressed gas. 

\subsection{Migration of solid matter and formation of rings}

\paragraph{}A certain period of time after the standing wave was established, most of the dust would have been concentrated in the regions of spherical antinodes. Due to increased dust concentration, the number of particle collisions significantly increased, causing their coalescence and increase in size and mass.  The gravitational field within the antinodes and the centrifugal force together caused dust to concentrate and form equatorial rings, with radii corresponding to the antinodes of the standing sound wave (Figure 4). 

\begin{figure}[ht]
\centering
\includegraphics[width=0.6\textwidth]{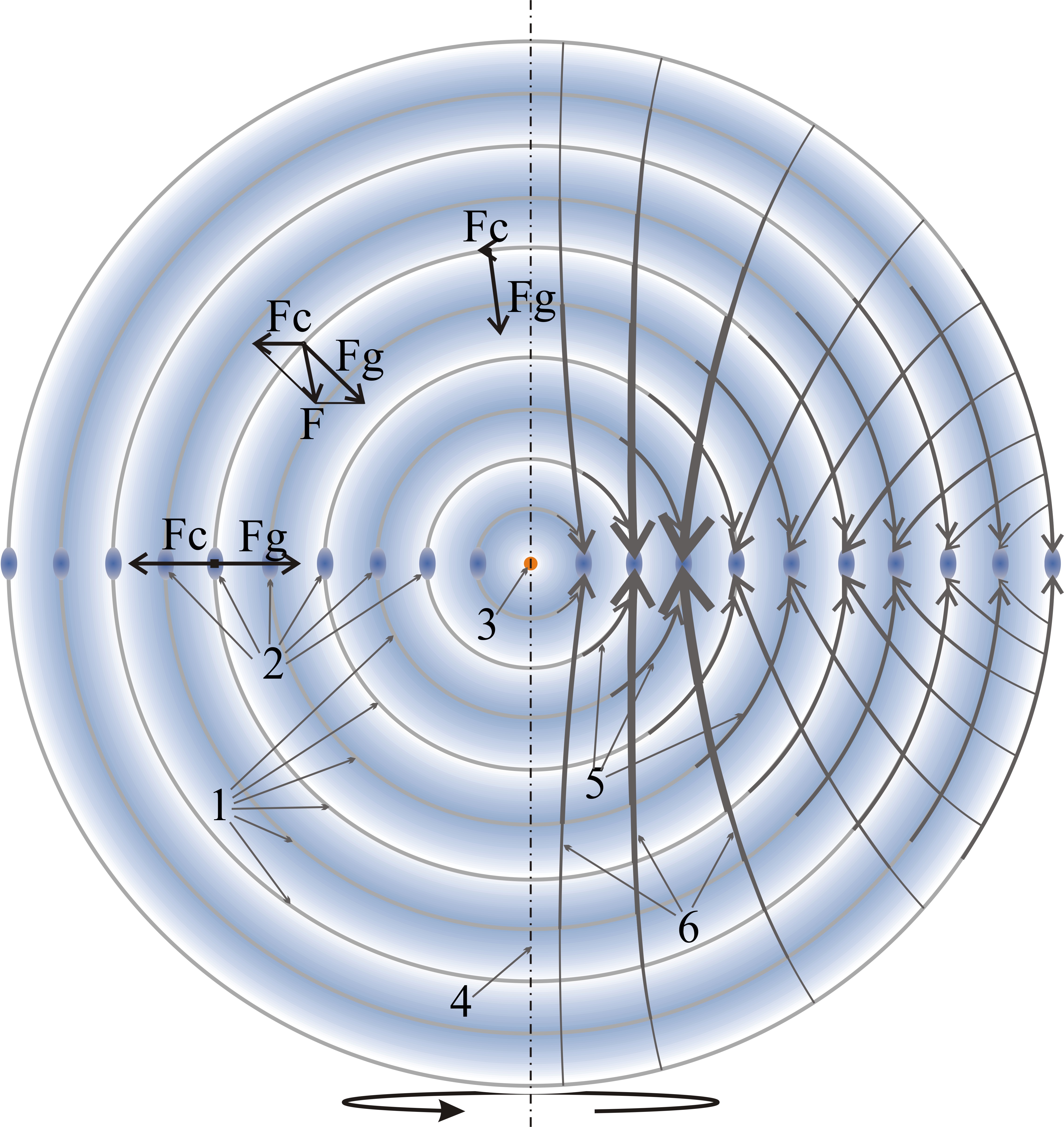}
\small \caption{Migration of solid matter in the spherical nebula. Fg represents the gravitational pull toward the centre and Fc represents the centrifugal force. (1) spherical dust clusters in the antinodes, (2) dust clusters in the equatorial rings, (3) protosun, (4) nebula rotational axis, (5) direction of dust migration in the antinodes, (6) direction of massive dust cluster migration across the antinodes under the influence of the resultant force F.}
\end{figure}

The model predicts that the increase in mass of the solid particles clusters also makes it increasingly difficult for them to stay at the antinodes of the standing sound wave as at the same time that they are affected by attraction from the central gas masses, principally the protosun. The largest clusters of dust fall toward the centre; this results in migration of solid matter from the outer regions of the pre-solar nebula towards the internal zones, while the centrifugal force does not allow the clusters to fall directly onto the protosun. The gravitational attraction of the gas masses in the antinodes directs the matter falling towards the spinning rings forming in the equatorial plane, where the solid matter comes to rest. A significant amount of solid matter collected in the region of the orbits of Venus and Earth, as it was concentrated here from most of the pre-solar nebula volume.

This dust, gathered in compact rings, is thousands of times more concentrated here than in the primary gas and dust nebula, greatly accelerating the coalescence of particles and leading to the emergence of increasingly massive dust clusters.

Agglomerated clusters of dust in narrow spinning rings move in almost identical circular orbits around the protosun and have very small relative velocities. Since their collisions do not result in fragmentation, they form massive planetesimals relatively quickly, which in turn agglomerate to form planets and their satellites. In this model, are no catastrophic collisions between the nascent planets, their satellites or other massive objects; the planets keep the circular orbits and planes of the equatorial dust rings. 

\subsection{Birth of the Sun}

\paragraph{}Gravitational contraction of the protosun, which had been significantly slowed as a result of regular explosions in the centre, still continued, and millions of years after the establishment of the standing sound wave gas temperature at the centre of the protosun became sufficiently high for thermonuclear fusion of hydrogen into helium to continue between explosions, resulting in the birth of a new star -- the Sun. The newborn Sun stopped augmenting the sound wave by periodic explosions, and the standing wave diminished. The gas shell previously supported by acoustic pressure began to shrink, forming gas-giant planets around already existing solid nuclei. Jupiter acquired the greatest share of mass, as it was located in the region of highest gas density. All the other gas planets were situated in a lower density environment and thus obtained smaller masses. Jupiter also received large amounts of gas from the inner-planet region, blown outwards by the strong solar wind in the first few millions of years after the birth of the Sun.

\subsection{The Oort cloud}

\paragraph{}The expanding shell that separated from the spherical part of the nebula during the passage of the wave front from the first explosion (see Figure 1) moved faster than the escape velocity $\nu_e$ and could not return to the centre, as the pull of gravity from the central masses was too weak at such distance. With its expansion, this shell accumulated increasing quantities of highly rarefied gas from the primary nebula at its edge, while its expansion rate gradually slowed and eventually stopped at a distance of about 1 light year from the centre. This formed a giant spherical region containing dust, ice and frozen gases particles in addition to gaseous hydrogen and helium. Over time, the gas component of the shell dissipated in space, while solid particles were concentrated in increasingly large chunks of ice and dust -- the cometary bodies -- to form the Oort cloud \cite{oort1950}  which has a weak gravitational connection with the central part of the system. 

\subsection{Neptune, the Kuiper belt and scattered-disc objects}

\paragraph{}After expansion of the gas shell ceased in the region of the Oort cloud, a weak reverse wave formed within the shell and moved towards the central spherical nebula of gas, where planetary nuclei had already emerged. Hundreds of thousands of years later the reverse wave from the Oort cloud collided with the outer boundary of the central spherical nebula, causing a redistribution of matter on the edge of the Solar System. Large gas and ice masses from the outer antinodes of the standing sound wave were shifted, and the planet Neptune formed a little closer to the Sun relative the original position of the 9th antinode, while its mass increased several times by capture of gas and ice from the reverse wave from the Oort cloud and from the outer antinodes. Scattered-disc objects such as the minor planet Eris acquired highly elongated orbits, as there was a long period under the influence of gravitational pull from the gas masses transported by the reverse wave from the Oort cloud. Minor planets in the Kuiper belt also gained significant eccentricity. The orbits of the low-mass objects in the scattered disc region and in the Kuiper belt have been changed so dramatically, that they had become to reach Neptune and Uranus, and under the influence of their gravity even began to go into the central region of the solar system and possibly caused the late heavy bombardment of the inner planets, in which was also attended cometary bodies from the Oort cloud, that have had unstable orbits in the first time after the formation.

The periphery processes were very slow, developing over many millions of years, and were relatively weak in their effect on the central region of the pre-solar system. The reverse wave from the Oort cloud only caused the formation of the Kuiper belt, an offset of Neptune's formation and its mass increase. The rest of the Solar System was and is still affected by the region of the Oort cloud only through comets.

\section{Discussion}

\subsection{Terrestrial and giant planets}

\paragraph{}The difference in chemical composition of the inner and outer planets is explained in our model by the migration of solid matter from the outer spherical dust shells to the internal ones (see Figure 4). Before concentrating in stable spinning dust rings, the heavy chemical elements, such as iron and silicate travelled farther towards the centre of the system and were incorporated by the inner planets, with Earth and Venus having more matter, and Mercury and Mars having less. The composition of the closest to the Sun Mercury at the same time has obtained a significant amount of iron. The region between Mars and Jupiter had insufficient solid matter left for a proper planet, so only the minor planets of the asteroid belt were formed. Lighter and more volatile chemical compounds, such as water, methane or ammonia remained in significant quantities in the colder regions beyond the asteroid belt, and became the compositional basis of the giant planets.

\subsection{Angular momentum distribution and dust structure}

\paragraph{}If we assume that the first powerful explosion in the centre of the protosun was not absolutely symmetrical, slight asymmetry in the explosion led to a redistribution of the angular momentum in the pre-solar nebula, with the less massive peripheral zone beginning to rotate faster. In addition, an initial asymmetry could produce the observed 7$^\circ$ equatorial plane tilt relative to the Sun's plane of rotation.

\begin{figure}[ht]
\centering
\includegraphics[width=0.55\textwidth]{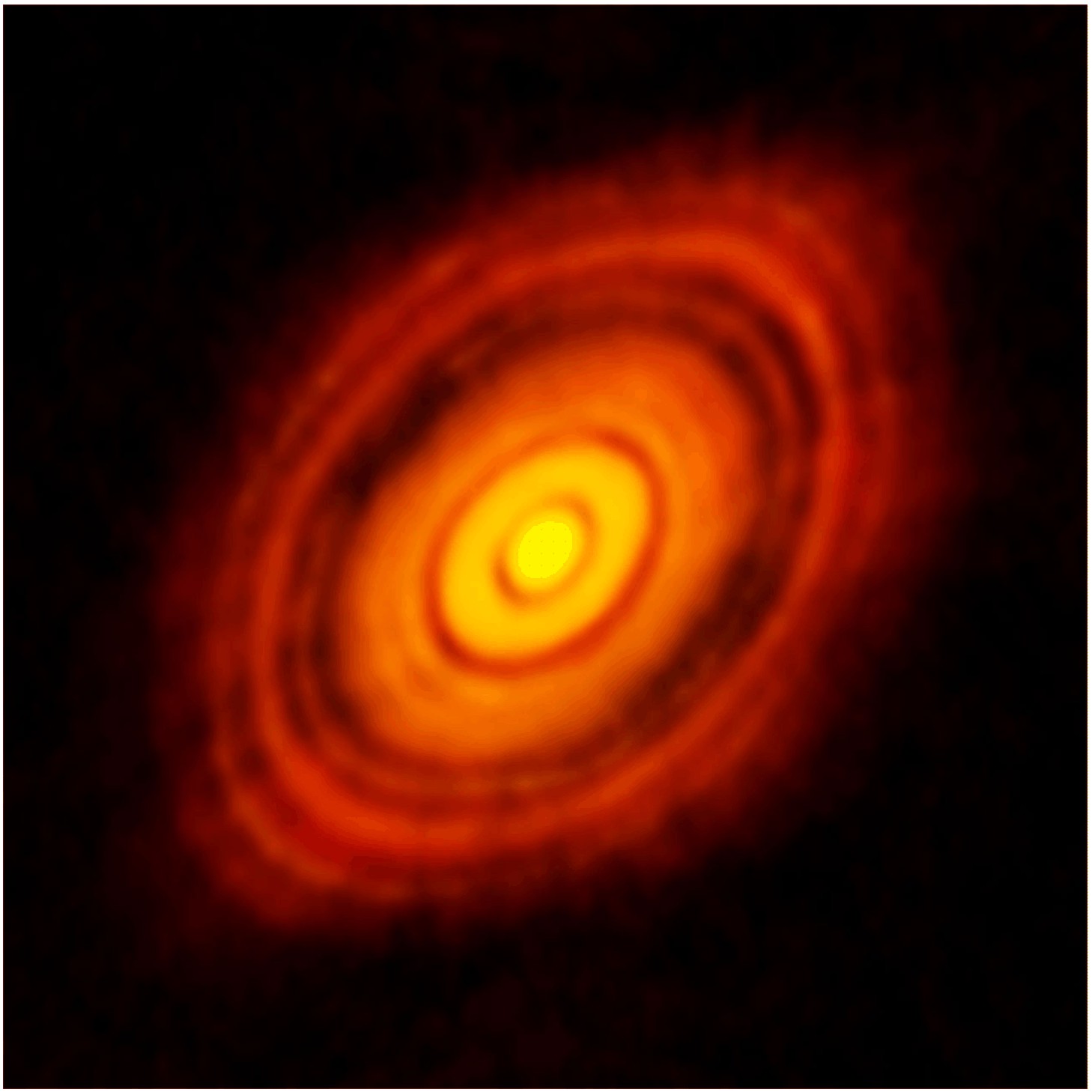}
\caption{An image of a protoplanetary dust nebula around HL Tauri received by ALMA (The Atacama Large Millimeter/submillimeter Array). Credit: ALMA (ESO/NAOJ/NRAO).}
\end{figure}

Currently the Sun has more than 99.8\% of the entire Solar System mass but only about 0.5\% of the total angular momentum. If the rotational kinetic energy were evenly distributed over the original pre-solar nebula, the resulting rotation (taking into account the pirouette effect) would be thousands or millions of times slower than the current rotation of the planets. The angular momentum was initially distributed evenly in the pre-solar nebula, and this means that such weak rotation cannot result in the process of dust disc formation as described by the nebular hypothesis. 

However, these discs are visible in images of some young stars (Figure 5), and in some cases their internal structure can be identified. If a spherical dust cluster such as described in this model were observed from one side, it would be impossible to observe any individual spheres as they would overlap in the line of sight. 

\begin{figure}[ht]
\centering
\includegraphics[width=0.55\textwidth]{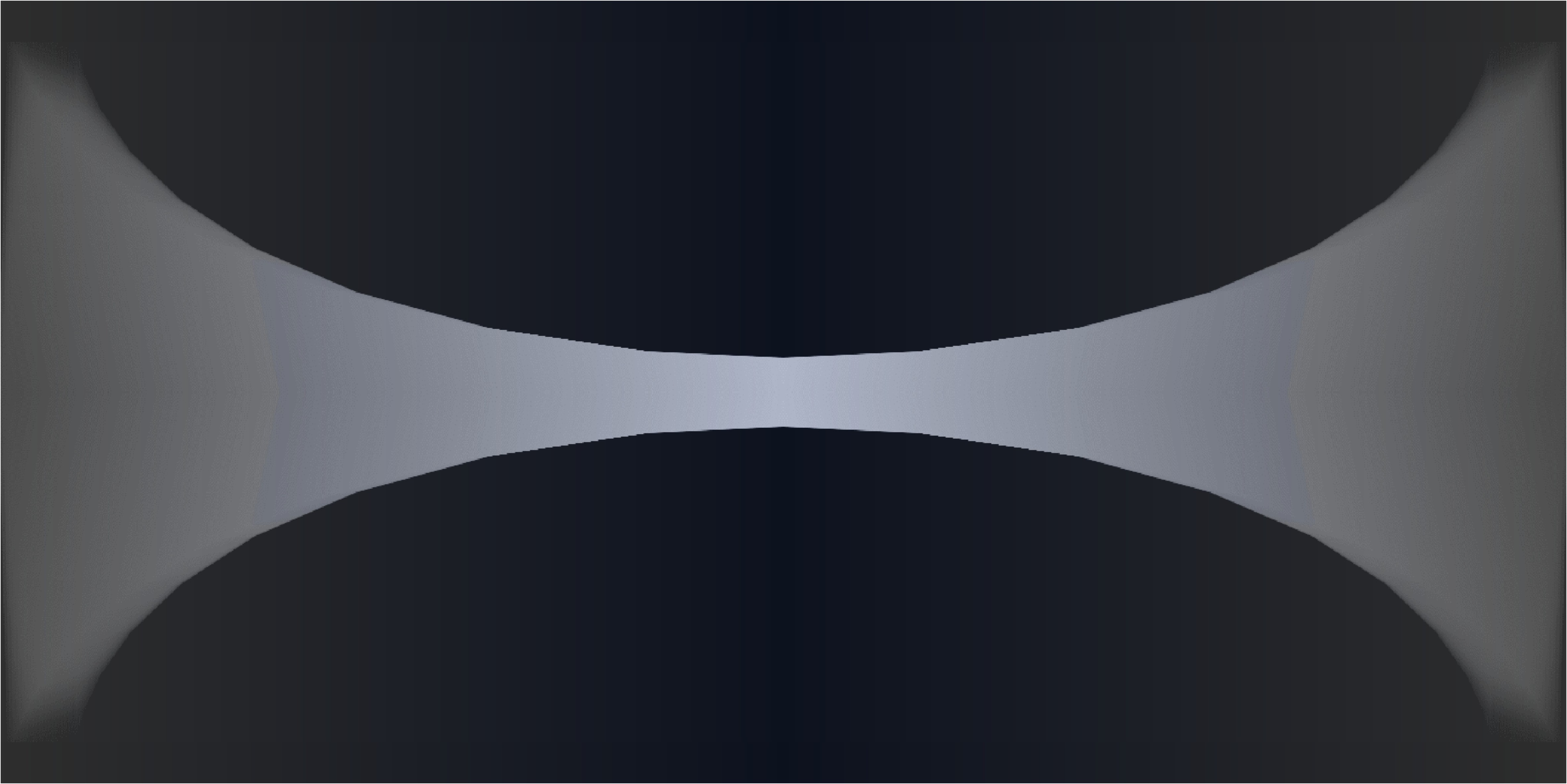}
\caption{The averaged in the line of sight distribution of dust in the spherical antinodes. Closer to centre the dust has already reached the equatorial plane, on the periphery it need to pass a much greater way, so formed a cup-shaped dust distribution structure around the protostar.}
\end{figure}

The stage at which a significant amount of dust has already accumulated in the equatorial plane would give such a system the appearance of a disc (Figure 6), despite the fact that there are significant gaps between individual dust clusters. 

\begin{figure}[ht]
\centering
\includegraphics[width=0.9\textwidth]{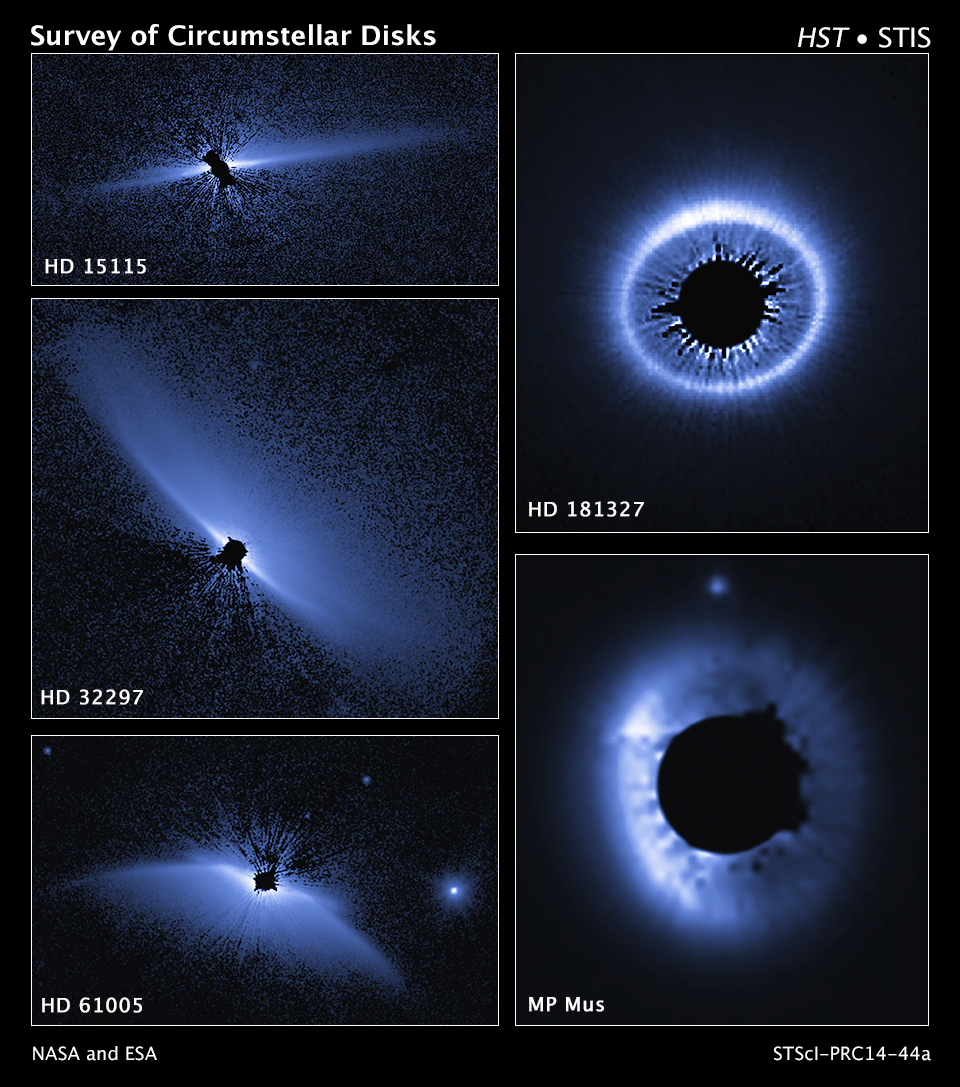}
\caption{The images from a NASA Hubble Space Telescope visible-light survey of the architecture of dust systems around young stars. Credit: STScI/Hubble, NASA \& ESA.}
\end{figure}

Figure 7 shows images from the NASA Hubble Space Telescope, showing the complex three-dimensional structure of the dust around young stars that are similar to the structure, which should be observed according to the proposed model (for more images see also Schneider et al. \cite{schn2014}). On these images we see the light reflected from the inner part the near side of the cup-shaped dust structure. In contrast to the images in visible light the image of HL Tauri on the Figure 5 received in submillimetre radiation can show the ring structures of dust around the less than 100000 year’s young star.

\subsection{Exoplanetary systems}

\paragraph{}Generally speaking, the explosion in a protostellar nebula centre according to our model can either accelerate the gas shell rotation around the future star, slow it down or even give it a spin in the opposite direction. The latter case has been observed in some exoplanetary systems, e.g. \cite{nari2009}-\cite{hebr2011}, and the standard nebular hypothesis cannot explain it, since it implies that the central star and its planets should always rotate and revolve in the same direction, following the rotation of the protostellar nebula. 

In this model, an asymmetry in the initial explosion in the centre of the protostar can also lead to a very strong tilt of the equatorial plane of the planetary system, with tilts of 45$^\circ$ or even 90$^\circ$ not impossible, e.g. \cite{pont2010}-\cite{hira2011}. 

Based on the data from the Kepler mission McQuillan et al. have measured the periods of the rotation of stars that have planets \cite{mcqu2013} and stars without planets \cite{mcqu2014}. Figure 8 shows a comparison of the distributions of rotation of these stars.

\begin{figure}[ht]
\centering
\includegraphics[width=0.7\textwidth]{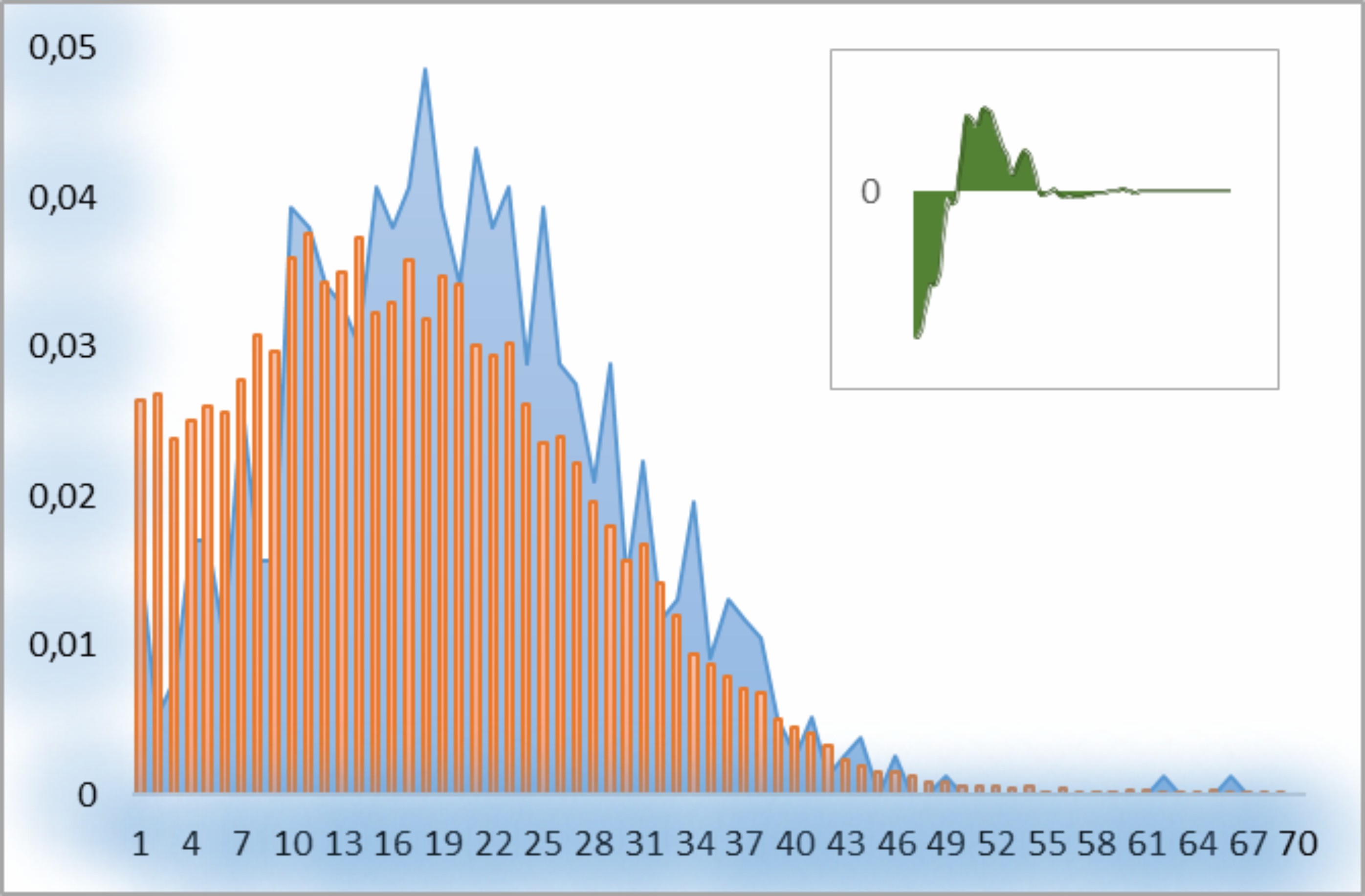}
\caption{The normalized distribution of rotation of stars with planets (blue) and without planets (orange). The horizontal axis represents the rotational period in days and the vertical axis the number of stars with periods in the range of one day, divided by the total number of stars (762 stars with planets and 34030 stars without planets). The inset shows the smoothed difference of these distributions.}
\end{figure}

One can see that the stars with planets rotate slower; according to our model these stars have slowed down its rotation as a result of an asymmetric explosion, which led to the formation of the planetary system. If the central protostar after the explosion accelerate its rotation and the rotation of the protostellar nebula slows down significantly or stops completely, the formation of planets is impossible and only a lone star would be formed from this protostellar nebula. It can also turn out that the nebula begins to rotate in the opposite direction, and in this case, the planets will have retrograde orbits.

In their work McQuillan et al. \cite{mcqu2013} showed that planets with a short orbital period are available only around slowly rotating stars and noted that the reason for this is unclear. The proposed model explains this fact by saying that closed to their stars planets, e.g. the so-called Hot Jupiter’s, were formed after a very strong asymmetric explosion, so the separation of the protostellar nebula (see Figure 1) has occurred in the immediate vicinity of the central protostar: the spherical nebula with a standing sound wave began spinning very rapidly and the central protostar at the same time began to rotate slowly.

\subsection{Background for a physical and mathematical model of the Solar System}

\paragraph{}Oscillations occur in an environment where gravitational field intensity and gas pressure change substantially within a wavelength. It can be assumed that the wavelength is inversely proportional to the product of average values of gravitational acceleration g, which is inversely proportional to the square of the distance to the centre of attraction, and gas pressure p, which in turn depends directly on g:
\begin{equation}
\lambda\sim\frac{1}{g \cdot p}
\end{equation}
or, supposing that $p \sim g$: 
\begin{equation}
\lambda \sim r^4,
\end{equation}

where $\lambda$ is the acoustic wavelength, g and p are the average values of gravitational acceleration and gas pressure as function of r, the distance from the centre. 

\begin{figure}[ht]
\centering
\includegraphics[width=0.7\textwidth]{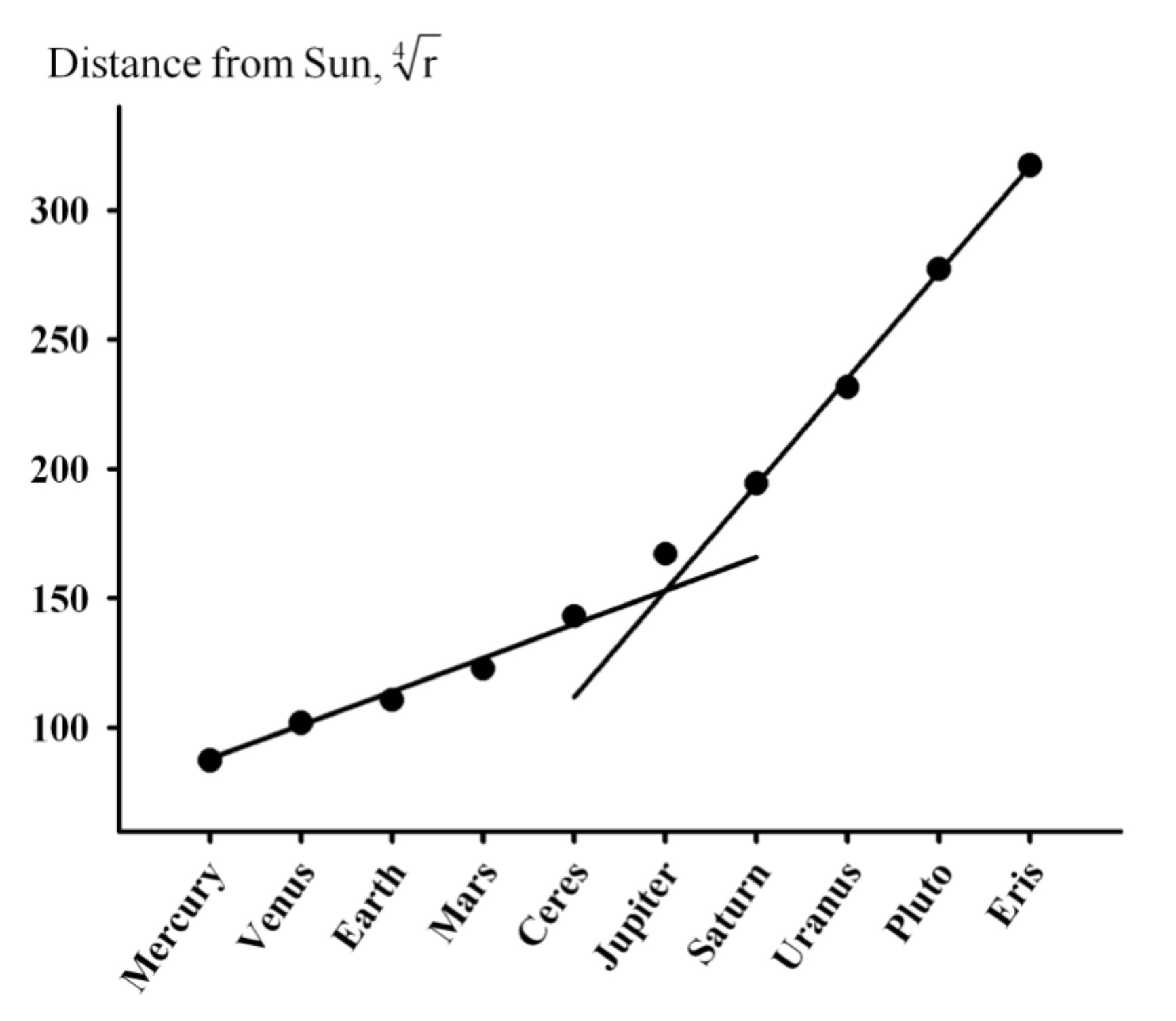}
\caption{Distribution of planetary distances from the Sun. The vertical axis shows the fourth root of distance in kilometres.}
\end{figure}

Figure 9 shows the distribution of planetary distances from the Sun. The planet Neptune is not shown on the figure for the reasons outlined in Section [2.6], its place being taken by Pluto, the main asteroid belt and the scattered disk are represented by the minor planets Ceres and Eris. The planets originated in the antinodes, so the figure also shows the distribution of the standing wave in the pre-solar gas nebula. We can see that the points on the graph may be approximated by two straight lines, supporting our hypothesis that locations of the standing wave antinodes depend on $r^4$. Closer to the Sun, the constant of proportionality determining the wavelength increases, which may be explained by the fact that the speed of the sound wave increases with rising gas temperature: $c \sim \sqrt{T}$; consequently, with the frequency $f = \frac{c}{\lambda}$ of the sound wave constant, 
\begin{equation}
\lambda (T) \sim \sqrt{T},
\end{equation}
where c is the speed of the sound wave, $\lambda$ is its wavelength and T is the temperature.
The dependence hypothesis $\lambda \sim r^4$ provides a background for developing a more complex physical and mathematical model for the Solar System’s origin, as well as for planetary systems around other stars.

\section{Conclusions}

\paragraph{}According to the proposed model of planetary origin from the gas and dust nebula in the field of a standing sound wave, only single stars or multiple stars with large distances between individual components can have planetary systems. Systems with closely located multiple star components cannot sustain the reverse wave after the explosion in the centre of one of the stars causes further periodic explosions due to the motion of the stars, so that it will fade out, while a gas-dust nebula that is not stabilized by the field of a standing sound wave will relatively swiftly accrete to the central stars without any planet formation. However, we do not claim that all the planetary systems must originate on the proposed model. In some rare cases, the planets can be formed, for example, as a result of a close passage of two or more stars and gas-dust nebulae, so may occur planetary systems around closely located multiple stars, e.g. Orosz et al. \cite{oros2012}.

Jensen and Akeson \cite{jens2014} describe the case of a binary young stars with components separated by about 386 AU, each of which is surrounded by a cloud of dust, and the plane of rotation of these clouds are strongly tilted relative to each other. The classical theory cannot explain this tilt, the proposed model explains this because the explosions in the centre of each of protostars occur independently of each other, and the dust clouds get their individual tilts.

The classical theory suggests that rotation of the protostellar cloud is essential for the formation of planets around the central star. The proposed model does not require such an assumption. If a gas-dust nebula does not rotate, after a first asymmetric explosion the central protostar and the peripheral gas masses will gain equal angular momentum, but opposite in sign. The gas-dust nebula will start spinning and at the same time a standing sound wave will emerge, beginning the process of planet formation. The resulting planetary system will have the central star and planets rotating in opposite directions, planetary orbits will be retrograde. 

The model predicts that the planetary systems with retrograde orbits can also be formed from a rotating protoplanetary nebula, when after an asymmetric explosion (i)the spherical nebula or (ii)the central protostar starts to rotate in opposite direction. In the first case, the planets will have wide orbits around rapidly rotating stars. In the second case, will be formed planets with short orbital periods, such as Hot Jupiter’s, around slowly rotating stars.

The Titius-Bode relation is not based on any physical laws and is only a mathematical formula to a series of numbers. Despite this, the fact that a simple formula with good accuracy describes the actual distances of the planets from the sun, suggesting, that behind it are standing certain physical regularities. Bovaird and al. \cite{bova2013,bova2015} have analysed a large number of exoplanet systems, consisting of at least from 4 planets, and found that in many of them distances of the planets from the centre correspond to the generalized Titius-Bode relation. The authors predicted the existence of 141 new planets in these systems. Huang and Bakos \cite{huan2014} performed an extensive search in the Kepler data for 97 of the predicted planets in 56 systems and have confirmed five of predictions. Our model provides a physical explanation of the distances of the planets from the central stars and we are confident that it will be able to more accurately predict the new planet in the already open exoplanetary systems. For this purpose, is necessary to develop a mathematical model of a standing wave in the protoplanetary nebula, which is our future target. 

A theory of the formation of the planetary systems must explain all or at least most of the known facts, but many of the current models consider only some of the facts, leaving all other ignored. There is, for example, abundant recent literature concerning transport of mass and angular momentum in turbulent, magnetized, differentially rotating disks. Measurements made in November 2014 by Rosetta spacecraft and Philae lander on comet 67P/Churyumov-Gerasimenko shown that the comet’s nucleus is not magnetised (e.g. Auster et al. \cite{aust2015}. This suggests that the magnetic field was absent at the stage of growth of the cometary nucleus and magnetism do not have played a role in the formation of the Solar System. From measurements of the gravitational field P\"atzold et al. \cite{patz2016} found that the comet 67P/Churyumov-Gerasimenko is low-density, highly porous (72-74 per cent) and homogeneous dusty body without large internal cavities. Massironi et al. \cite{mass2015} showed that the comet was merged from two kilometer-sized objects collided at very low relative speeds. This implies that the comet was formed from particles, coalesced at very low relative velocities, and never experienced deforming its structure collisions with other bodies. This in turn means that the formation of the solar system did not occur under the conditions of turbulence and chaos, as described in many models.

The proposed scenario of the planetary system origin answers many open questions related to the origin and evolution of the Solar System. Several known facts and their explanations within the framework of this model are listed below:

\paragraph{}\emph{4.1 Planetary distances to the Sun are not random -- there are certain regularities (Titius-Bode law).}

-- Planets are formed in the antinodes of a giant standing sound wave, emerging after a powerful thermonuclear explosion in the centre of the protosun and repeated passage of forward and backward sound waves through the spherical protoplanetary gas-dust nebula.

\paragraph{}\emph{4.2 There are internal silicate planets and outer gas giant planets. The hypothesis of a rotating protoplanetary disk cannot explain such distribution, as the rotating disk has the whole mass of dust influenced by the centrifugal force, which prevents migration of the matter.}

-- Distinction between inner and outer planets is explained by migration of the solid matter from the spherical dust concentration zones in the antinodes of the standing wave.

\paragraph{}\emph{4.3 The Sun contains 99.8\% of the mass, but only 0.5\% of the angular momentum of the Solar System.}

-- Asymmetry of the first explosion in the protosun centre resulted to redistribution of angular momentum: the rotation of the peripheral portion of the gas-dust nebula was significantly accelerated, acquiring an increment of angular momentum from the large gas mass emitted from the protosun during explosion.

\paragraph{}\emph{4.4 There is a 7$^\circ$ tilt of the Sun equatorial plane in relation to the average plane of the planetary orbits.}

-- The equatorial plane of the planetary system tilted during the first powerful explosion in the centre of the pre-solar nebula because of the small asymmetry of the explosion.

\paragraph{}\emph{4.5 The Oort cloud is a source of comets visiting the inner Solar System. Its existence is not confirmed by direct observations, but is very likely.}

-- The expanding shell that separated from the interior part of the pre-solar nebula during the passage of the shock wave from the first explosion concentrates large masses of rarefied gas in front of it and stops at a distance of about 1 light year from the Sun, forming a spherical region where comets are formed – the Oort cloud. 

\paragraph{}\emph{4.6 Neptune is closer to the Sun than what is implied by the Titius-Bode distribution.}

-- Neptune was shifted towards the Sun by a backward wave from the Oort cloud.

\paragraph{}\emph{4.7 The mass of Neptune (17.5 M$_\oplus$) is significantly greater than what could be expected based on the decreasing sequence of giant planet masses: Jupiter (318 M$_\oplus$), Saturn (95.3 M$_\oplus$) and Uranus (14.5 M$_\oplus$).}

-- Neptune gained a significant (a number of times) mass increase from the backward wave of the Oort cloud.

\paragraph{}\emph{4.8 The Kuiper Belt contains unexpectedly high quantities of wide binary objects (e.g. Petit et al. \cite{peti2008}, Parker et al. \cite{park2010}).}

-- The growing clumps of matter in rotating rings of dust had very small relative velocities and formed of them in the Kuiper belt large objects approaching at low speeds form gravitationally bound wide binary systems. In our model, the planets retain their orbits unchanged throughout all time of existence solar system; therefore binaries in the Kuiper belt were not destroyed by any approaching massive bodies.

\paragraph{Acknowledgments:} This research has made use of the NASA Exoplanet Archive, which is operated by the California Institute of Technology, under contract with the National Aeronautics and Space Administration under the Exoplanet Exploration Program.

\section*{Appendix A}
Currently there are no detailed models of the evolution of protostars in the period from the beginning of thermonuclear reactions to sustainable hydrogen burning and the formation of convective energy transfer zone. When in the process of gravitational contraction temperature in the center of the protostar reaches 10 million Kelvin, start the fusion reactions of hydrogen into helium. These reactions, however, are very slow with the characteristic time millions of years, so a relatively small amount of released energy cannot stop the gravitational contraction of a protostar and the temperature in the central region continues to rise. With an increase in temperature increases the rate of fusion reactions and the amount of released energy, and at some point the contraction of the protostar’s centre stops. The central region of the protostar at this time is in a very unstable state; the hydrogen burning begins to occur explosively and from the centre to the periphery begins to spread the shock wave. After the first explosion the central region expands and cools, the rapid hydrogen burning stops, then the compression process is repeated and the next explosion occurs. If a standing sound wave arises in the protostellar nebula, the formation of a planetary system begins, as described in our model. If the standing wave does not appear, the explosions would continue regularly or irregularly until a stable burning of hydrogen be established, at the same time from the surface of the protostar can occur a significant ejection of matter, which in some cases can be observed as a jets, outflows or outbursts, e.g. \cite{ball1995}-\cite{stan2003}.

During the passage of the shock wave from the first explosion through the central region of protostar are produced so large pressures and temperature that at the wave front can begin nucleosynthesis of elements heavier than helium, including the elements heavier than iron. Authors, who examined the live in the early solar system short-lived isotopes such as, for example, $^{26}$Al, $^{41}$Ca (e.g. \cite{boss1998,meyer2000}) or $^{247}$Cm \cite{tiss2016}, presume that they were injected in the solar system by supernova explosions within a very short, not more than a few millions of years, time before the formation of the solar system. This, in turn, requires the existence of some particular conditions for the emergence of the solar system. Our model eliminates the need for any special preconditions and does not put the solar system in a unique position among other planetary systems, because the short-lived isotopes can be produced during the formation of the sun.

\end{document}